\documentclass[fleqn,usenatbib,useAMS]{mnras}

\usepackage{newtxtext,newtxmath}
\usepackage{natbib}
\usepackage{xspace}
\usepackage{multirow}
\usepackage{dcolumn}
\usepackage{mathrsfs}

\usepackage[T1]{fontenc}

\DeclareRobustCommand{\VAN}[3]{#2}
\let\VANthebibliography\thebibliography
\def\thebibliography{\DeclareRobustCommand{\VAN}[3]{##3}\VANthebibliography}

\usepackage{graphicx}	
\usepackage{amsmath}	
\usepackage{multicol}        
\usepackage{bm}		
\usepackage{pdflscape, verbatim}

\title{Kinematic Constraints on Spatial Curvature from Supernovae Ia and Cosmic Chronometers}

\author[J. F. Jesus, R. Valentim, P.H.R.S. Moraes, M. Malheiro]{
J. F. Jesus,$^{1,2}$\thanks{E-mail: jf.jesus@unesp.br}
R. Valentim,$^{3}$\thanks{E-mail: valentim.rodolfo@unifesp.br}
P.H.R.S. Moraes$^{4}$\thanks{E-mmail: moraes.phrs@gmail.com}
M. Malheiro$^{5}$\thanks{E-mail: malheiro@ita.br}
\\
$^{1}$Universidade Estadual Paulista (UNESP), Campus Experimental de Itapeva - R. Geraldo Alckmin 519, 18409-010, Itapeva, SP, Brazil\\
$^{2}$Universidade Estadual Paulista (UNESP), Faculdade de Engenharia, Guaratinguet\'a, Departamento de F\'{\i}sica e Qu\'{\i}imica - Av. Dr. Ariberto \\Pereira da Cunha 333, 12516-410, Guaratinguet\'a, SP, Brazil\\
$^{3}$Departamento de F\'{\i}sica, Instituto de Ci\^encias Ambientais, Qu\'{\i}micas e Farmac\^euticas (ICAQF), Universidade Federal de S\~ao Paulo - UNIFESP, \\Rua S\~ao Nicolau no.210, Centro, 09913-030, Diadema - SP, Brazil\\
$^{4}$Universidade de S\~ao Paulo, Instituto de Astronomia, Geof\'isica e Ci\^encias Atmosf\'ericas, Rua do Mat\~ao 1226, 05508-090 S\~ao Paulo, SP, Brazil\\
$^5$Departamento de F\'{\i}sica, Instituto Tecnol\'ogico de Aeron\'autica (ITA), 12228-900, S\~ao Jos\'e dos Campos - SP, Brazil
}

\date{Accepted XXX. Received YYY; in original form ZZZ}

\pubyear{2015}

\DeclareMathOperator{\atan}{arctan}

\newcommand{\be}{\begin{equation}}
\newcommand{\ee}{\end{equation}}
\newcommand{\bea}{\begin{eqnarray}}
\newcommand{\eea}{\end{eqnarray}}
\newcommand{\texpdf}{\texorpdfstring}
\newcommand{\like}{\mathscr{L}}

\newcommand{\sinn}{\,\mathrm{sinn}\,}

\begin{document}
\label{firstpage}
\pagerange{\pageref{firstpage}--\pageref{lastpage}}
\maketitle

\begin{abstract}
An approach to estimate the spatial curvature $\Omega_k$ from data independently of dynamical models is suggested, through kinematic parameterizations of the comoving distance ($D_{C}(z)$) with third degree polynomial, of the Hubble parameter ($H(z)$) with a second degree polynomial and of the deceleration parameter ($q(z)$) with first order polynomial. All these parameterizations were done as function of redshift $z$. We used SNe Ia dataset from Pantheon compilation with 1048 distance moduli estimated in the range $0.01<z<2.3$ with systematic and statistical errors and a compilation of 31 $H(z)$ data estimated from cosmic chronometers. The spatial curvature found for $D_C(z)$ parametrization was $\Omega_{k}=-0.03^{+0.24+0.56}_{-0.30-0.53}$. The parametrization for deceleration parameter $q(z)$ resulted in $\Omega_{k}=-0.08^{+0.21+0.54}_{-0.27-0.45}$. The $H(z)$ parametrization has shown incompatibilities between $H(z)$ and SNe Ia data constraints, so these analyses were not combined. The $D_C(z)$ and $q(z)$ parametrizations are compatible with the spatially flat Universe as predicted by many inflation models and data from CMB.
This type of analysis is very  appealing as it avoids any bias because it does not depend on assumptions about the matter content of the Universe for estimating $\Omega_k$.
\end{abstract}




\section{Introduction}

The evidence that the universe is accelerated first came from Supernovae Ia (SNe Ia) observations \citep{SN1,SN2,SN3,SN4,SN5,union,union2,union21} and was subsequently complemented by data from Cosmic Microwave Background (CMB) radiation \citep{WMAP1,WMAP2,planck}, Baryonic Acoustic Oscillations (BAO) \citep{BAO1,BAO2,BAO3,BAO4,BAO5}, and the Hubble parameter $H(z)$ data \citep{Omer,Omer2,Omer3}. The acceleration phase of the universe can be supported by a simple theoretical model using the cosmological constant $\Lambda$ plus Cold Dark Matter component \citep{reviewDM,bookDM,weinberg2013}. This model has cosmological parameters that have been restricted more and more, and have become very precise by observational data \citep{planck,Omer2,sharov}.

In addition to the ``standard'' model that emerges from $\Lambda$ in the context of Cold Dark Matter, other models have been proposed to explain the problem of accelerated expansion of the universe. Many of these models have as their main idea, a dark energy fluid that produces a negative pressure that would fill the universe \citep{peebles,sahni2}. Many hypotheses suggest the nature of this unknown fluid as scalar fields and quintessential models \citep{amendola2000,sahni2000,chiba2000,capozziello2002, khurshudyan2014}. Other approaches dealing with accelerated expansion come from modified gravity theories \citep{volkov}, $f(R)$ and $f(T)$, with $R$ and $T$ being the Ricci and Energy-Momentum trace scalars \citep{Moraes2019, Harko2011}, respectively,  which generalize the general theory of relativity \citep{fR,fR3,fR4}, are also investigated; models based on extra dimensions: as models of the braneworld \citep{randall, pomarol, langlois, shir, cline},
strings \citep{string} and Kaluza-Klein \citep{klein}, among other works. Having adopted a specific model, cosmological parameters can be determined based on statistical analysis of observational data. All of these suggested hypotheses need first of all to be sifted through observational data. This is the way to study cosmology in the present times.

On the other hand, some works attempt to investigate the history of the universe independently of dynamical models. These approaches are called cosmography models or cosmokinetic models \citep{kine1,kine2,kine3,kine4,kine5,kine6}. The work of \citep{CapozzielloEtAl2020} suggests comparing two different parameterizations: auxiliary variables versus Pad\'e polynomials for high redshifts. Both approaches are made in the context of cosmography, where the scale parameter $a$ is expanded on Taylor series at the present time $t_0$. This work compares both analysis through the AIC (Akaike Information Criterion) and the BIC (Bayesian Information Criterion) and showed that the parameterization from the Pad\'e expansion was more promising in the estimate of $H_0$, $q$ and $j$.

In this paper, we will refer to them only as kinematic models, whose name comes from the idea that the universe expansion (or its kinematics) is described by the Hubble expansion rate $ H = \frac{\dot{a}}{a}$, deceleration parameter $q=-\ddot{a}a/\dot{a}^2$ and the jerk parameter $j = -\dddot{a}a^3/(a\dot{a}^3)$, where  $a$ is the Friedmann-Lema\^itre-Robertson-Walker (FLRW) metric scale factor. That is, this approach relies only on the Cosmological Principle, which states that the Universe is statistically homogeneous and isotropic at large scales. Assuming an FLRW metric, which is exactly homogeneous and isotropic, one then looks for hints of $a(t)$ evolution directly from data. In this parameterization, dark matter dominated the $q=-1/2$ universe, while the $\Lambda$CDM accelerated model has $j=-1$. These analyses allow us to study the transition from decelerated to accelerated phases, while the $j$ parameter allows you to study deviations from the cosmic concordance model without the restriction of a specific model.

There are several works in the literature that estimated cosmological parameters independently of energy content, in which some authors used parameterization in these estimates \citep{MortsellJonsson11,YuWang16}. In \citep{SaponeEtAl14}, an expansion of the comoving distance was made, as a function of $\Omega_k$, while \citep{HuillierShafieloo16} reconstructed the luminosity distance with a lognormal kernel using data from BAO and SNeIa (BOSS DROSS 12 and JLA). Furthermore, in \citep{WeiWu16} the distance modulus $\mu_H(z)$ was reconstructed from $H(z)$ data using Gaussian processes and compared with $\mu_{\text{SN}}(z)$ from SNe Ia to estimate $\Omega_k$. In \citep{HeavensEtAl14}, $\Omega_k$ is estimated from BAO data regardless of the model. Other parameters were used by \citep{MontanariRasanen17} to analyse the consistency conditions of FRW. And other authors, such as \citep{CollettEtAl19} and \citep{LiaoEtAl19} estimated the values of $H_0$ and $\Omega_{k}$ from gravity lensing data and SNe Ia data where the latter used Gaussian Processes (GP).

All these parametrizations help to reconstruct the Universe evolution without mentioning the dynamics, that is, without the use of Einstein's Equations. Furthermore, by using the FLRW metric geometry, we may relate these parametrizations ($H(z)$, $q(z)$) to spatial curvature and cosmological distances: luminosity-distance ($d_L$) and angular diameter distance ($d_A$). So, by using distance data, like the ones provided by SNe Ia, one may constrain spatial curvature, without assuming any particular Cosmology dynamics. This was first shown by \citep{ClarksonEtAl08}.

A first test of this method was done by M\"ortsell and Clarkson \citep{MortClark09}. By using only SNe Ia data and 3 parametrizations of $q(z)$, namely, constant, piecewise and linear on $a$, they have shown that the Universe is currently accelerating regardless of spatial curvature, but could not conclude about an early expansion deceleration. By combining SNe Ia data with BAO, they concluded that the Universe could have early deceleration only for a flat or open Universe ($\Omega_k\geq0$). It has been shown that future 21 cm intensity experiments can improve model-independent determinations of the spatial curvature \citep{WitzemannEtAl18}.

\citep{YuEtAl18} have compiled 36 data of $H(z)$, where 31 are measured by using the chronometric technique, while 5 come from BAO (Baryon Acoustic Oscillations) observations. This work used Gaussian Processes (GP) to estimate the continuous function of $ H(z)$ with values of $H_{0}$, $z_{t}$ and $\Omega_{k}$ to test the $\Lambda$CDM model. They have found $H_0=67\pm4\mathrm{\,km\,s}^{-1}\mathrm{Mpc}^{-1}$. Using the profile of $H(z)$ function they estimate limits for the curvature parameter $\Omega_k$. It was found that the transition from deceleration to acceleration redshift is $0.33<z_t<1.0$ to $1\sigma$ of significance and the value of $\Omega_{k}=-0.03\pm0.11$, which is consistent with a spatially flat universe. \citep{DiValentino2019} argue that there is a crisis in Cosmology due to interval values of $\Omega_k$ obtained from Planck Legacy 2018 (PL2018), $-0.095<\Omega_k<-0.007$, incompatible with a spatially flat Universe, at more than 99\% c.l.

In the present work we study the spatial curvature by means of a third order parametrization of the comoving distance, a second order parametrization of $H(z)$ and a linear parametrization of $q(z)$. By combining luminosity distances from SNe Ia \citep{pantheon} and $H(z)$ measurements \citep{MaganaEtAl18}, it is possible to determine $\Omega_k$ values in these cosmological models, independently of the matter content of the Universe. In this type of approach, we obtain an interesting complementarity between the observational data and, consequently, tighter constraints on the parameter spaces.


The paper is organized as follows. In Section \ref{basic}, we present the basic equations concerning the obtainment of $\Omega_k$ from comoving distance, $H(z)$ and $q(z)$. Section \ref{samples} presents the dataset used and the analyses are presented in Section \ref{analysis}. Conclusions are left to Section \ref{conclusion}. 



\section{Basic equations}\label{basic}

For general cosmologies, the spatial curvature could not be constrained from a simple parametrization of the cosmological observables. However, as curvature relates to geometry, if one parametrizes the dynamics, the geometry can be constrained through the relation among distances and dynamic observables. To realize this, let us assume as a premise the validity of the Cosmological Principle, which leads us to the Friedmann-Lema\^itre-Robertson-Walker metric:
\be
ds^2=-dt^2+a(t)^2\left[\frac{dr^2}{1-kr^2}+r^2(d\theta^2+\sin^2\theta d\phi^2)\right].
\ee

In the context of the FLRW metric, the line-of-sight distance distance can be estimated. This is the distance between two objects in the universe that remain constant if the objects are moving with the Hubble flow  \citep{Hogg92}. The line-of-sight comoving distance between an object in redshift $z$ and us is given by
\be
d_C=d_H\int_0^z\frac{dz'}{E(z')},
\label{dCEz}
\ee
where $d_H\equiv\frac{c}{H_0}$ is the Hubble distance and the dimensionless Hubble parameter  $E(z)\equiv\frac{H(z)}{H_0}$. As all cosmological distances scale with $d_H$, we shall adopt the notation where a distance written in upper case ($D_i$) is dimensionless, while a distance written in lower case ($d_i$) is dimensionful and $d_i\equiv d_HD_i$. So, we may write
\begin{equation}
\label{Dc}
D_C(z) = \int_0^z \frac{dz'}{E(z')}.
\end{equation}

From this we may obtain the transverse comoving distance. The comoving distance between two events at the same redshift but separated on the sky by some angle $\delta\theta$ is $d_M\delta\theta$ and the transverse comoving distance is related to the line-of-sight comoving distance as:
\be
d_M=d_H\left\{\begin{array}{ll}
     \frac{1}{\sqrt{\Omega_k}}\sinh\left[\sqrt{\Omega_k}D_C\right] & \mathrm{for}\,\,\Omega_k>0,\\
     D_C& \mathrm{for}\,\,\Omega_k=0,\\
     \frac{1}{\sqrt{-\Omega_k}}\sin\left[\sqrt{-\Omega_k}D_C\right] & \mathrm{for}\,\,\Omega_k<0.\\
    \end{array}\right.
\label{dM}
\ee
where we have used the curvature parameter density $\Omega_k\equiv-\frac{k}{a_0^2H_0^2}$. By defining the following function
\be
\sinn(x,\Omega_k)\equiv \left\{\begin{array}{ll}
     \frac{1}{\sqrt{\Omega_k}}\sinh\left[x\sqrt{\Omega_k}\right] & \mathrm{for}\,\,\Omega_k>0,\\
     x& \mathrm{for}\,\,\Omega_k=0,\\
     \frac{1}{\sqrt{-\Omega_k}}\sin\left[x\sqrt{-\Omega_k}\right] & \mathrm{for}\,\,\Omega_k<0,\\
    \end{array}\right.
\label{sinn}
\ee
Eq. (\ref{dM}) can be simplified as
\be
d_M=d_H\sinn(D_C,\Omega_k).
\label{dMsinn}
\ee

The luminosity distance $d_L$ is defined by the relationship between bolometric flux $S$ and bolometric luminosity $L$:
\be
d_L=\sqrt{\frac{L}{4\pi S}}.
\ee
We may relate it to the transverse comoving distance by
\begin{equation}
D_L(z)=(1+z)D_M(z).
\label{DL}
\end{equation}

We shall briefly mention the dynamics here just to show how the curvature density parameter definition emerges. As it is well known, the Friedmann equations can be written as:
\bea
H^2&=&\frac{8\pi G\rho_T}{3}-\frac{k}{a^2}\label{H2},\\
\frac{\ddot{a}}{a}=\dot{H}+H^2&=&-\frac{4\pi G}{3}(\rho_T+3p_T),
\label{acel}
\eea
where $\rho_T$ represents the total energy density and $p_T$ the total pressure. As it can be seen, the spatial curvature contributes to the Hubble parameter through Eq. (\ref{H2}), while it does not contribute to acceleration ($\ddot{a}$) explicitly (\ref{acel}). The  Friedmann equation shows that if we know the matter-energy content of the Universe, we can estimate its spatial curvature. This can be clearly seen if we rewrite Eq. (\ref{H2}) as
\be
\Omega_T+\Omega_k=1 \,,
\ee
where $\Omega_T\equiv\frac{8\pi G\rho_T}{3H^2}$ is the  total energy density parameter and $\Omega_k\equiv-\frac{k}{a^2H^2}$ is the curvature parameter.

Here, we intend to obtain constraints over spatial curvature without making any assumptions about the matter-energy content of the Universe. Thus, we shall assume kinematic expressions for the observables like $H(z)$, $q(z)$ and $D_C(z)$.

Assuming this kinematic approach, we can see that $H(z)$ data alone cannot constrain spatial curvature, but luminosity distances from SNe Ia can constrain it through the $\Omega_k$ dependence in Eq. (\ref{dM}). Concerning the deceleration parameter $q(z)$, it can be given as
\begin{equation}
 q(z)=-\frac{\ddot{a}}{aH^2}=\frac{1+z}{H}\frac{dH}{dz}-1\,, \label{qzH}
\end{equation}

So, as expected from Eq. (\ref{acel}), a $q(z)$ kinematical parametrization will not depend explicitly on spatial curvature, however, the spatial curvature can be constrained through the distance relation (\ref{dM}).

Therefore, we may access the value of $\Omega_k$ through a parametrization of both $q(z)$ and $H(z)$
. As a third method we can also parametrize the line-of-sight comoving distance, which is directly related to the luminosity distance, in order to obtain the spatial curvature. In what follows we present the three different methods considered here.

\subsection{Choice of parametric functions for \texpdf{$D_C(z)$}{Dc(z)}, \texpdf{$H(z)$}{H(z)} and \texpdf{$q(z)$}{q(z)}}
We shall perform a model selection in order to find the ideal polynomial that better describes the data for the $d_C$, $H$ and $q$ parametrizations. The best Bayesian tool for model selection is Bayesian Evidence \citep{Kass95,ElgMult06,GuimaraesEtAl09,JesusEtAl17}. However,
the Bayesian Evidence, in general, is given by 
multidimensional integrals over the parameters, so it is
 usually hard to evaluate. A way around this 
difficulty is by using its approximation, first obtained by 
Schwarz \citep{Schwarz78,Liddle04}, known as 
BIC. Bayesian Information Criterion (BIC) 
\citep{ValentimEtAl11,ValentimEtAl11b,SzydlowskiEtAl15} heavily penalizes models with different number of free parameters.

Here we use BIC (Bayesian Information Criterion) in order to find the ideal polynomial order in each of the parametrizations aiming to find model-independent constraints on spatial curvature. The BIC is given by:
\be
\text{BIC}=-2\ln\mathcal{L}_{max}+p\ln N \,,
\ee
where $p$ is the number of free parameters and $N$ is the number of data. As the likelihood is given by
\be
\mathcal{L}=e^{-\frac{\chi^2}{2}}\,,
\ee
then, we may write
\be
\text{BIC}=\chi^2_{min}+p\ln N \,.
\ee

The interpretation of $\Delta$BIC outcomes is described in Table \ref{tab:bic}.


\begin{table}
   \centering   
   \setlength{\arrayrulewidth}{2\arrayrulewidth}  
   \begin{tabular}{ccccc} 
      \hline\hline
      $\Delta\mbox{BIC}$  & Support \\
      \hline
      $\Delta\mbox{BIC} \leq1$ &  No worth more than a bare mention \\
       $1\leq\Delta\mbox{BIC} \leq3$ &   Significant/Weak\\
      $3\leq\Delta\mbox{BIC} \leq5$ &  Strong to very strong/Significant\\
      $\Delta\mbox{BIC} >5$ &  Decisive/Strong \\
      \hline \hline
   \end{tabular}
   \caption{Bayesian Information Criterion}
   \label{tab:bic}
\end{table}

The results for the three parametrizations can be seen on Table \ref{tabBIC}.

\begin{table*}
\begin{tabular}{|c|c|c|c|c|c|}
\hline
\, Parametrization \, & \, Polynomial order \, & $\chi^2_{min}$ & $\chi^2_{red}$ & BIC & $\Delta$BIC\\ \hline
\multirow{4}{*}{$D_C(z)$} & 1 & \, 1250.825 \, & \, 1.16140 \, & \, 1264.793 \, & \, +193.915 \, \\ \cline{2-6}
& 2 & \, 1049.927 \, & \, 0.97577 \, & \, 1070.878 \, & \, 0 \, \\ \cline{2-6} 
                          & 3 & 1043.839 & 0.97101 & 1071.775 & +0.897\\ \cline{2-6} 
                          & 4 & 1042.471 & 0.97064 & 1077.390 & +6.512\\ \hline
\multirow{3}{*}{$H(z)$}   & 1 & 1059.386 & 0.98456 & 1080.338 & +9.253\\ \cline{2-6} 
                          & 2 & 1043.150 & 0.97037 & 1071.085 & 0\\ \cline{2-6} 
                          & 3 & 1042.975 & 0.97111 & 1077.894 & +6.809\\ \hline
\multirow{4}{*}{$q(z)$}   & 0 & 1066.636 & 0.99130 & 1087.588 & +16.036\\ \cline{2-6}
                          & 1 & 1043.617 & 0.97081 & 1071.552 & 0\\ \cline{2-6}
                          & 2 & 1042.919 & 0.97106 & 1077.838 & +6.286\\ \cline{2-6} 
                          & 3 & 1042.375 & 0.97146 & 1084.278 & +12.726\\ \hline
\end{tabular}
\caption{Bayesian model comparison for different parametrizations.}
\label{tabBIC}
\end{table*}

As can be seen on Table \ref{tabBIC}, the ideal polynomial order for $D_C(z)$, $H(z)$ and $q(z)$ are 2, 2 and 1, respectively. However, for $D_C(z)$, the third degree polynomial can not be discarded by this analysis ($\Delta\text{BIC}<1$). We have tested the second degree parametrization for $D_C(z)$ and have found a too close Universe ($\Omega_k=-0.49\pm0.14$ at 68\% c.l.), which was in disagreement with the $q(z)$ parametrizations and with other data, like CMB \citep{Planck18}. As the third order parametrization can not be discarded by this analysis, we chose to work with this order for $D_C(z)$.

\subsection{\texpdf{$\Omega_k$}{Wk} from line-of-sight comoving distance, \texpdf{$D_C(z)$}{Dc(z)}}

In order to put limits on $\Omega_k$ by considering the line-of-sight comoving distance, we can write $D_C(z)$ as a third degree polynomial such as: 
\begin{equation}
 D_C = z + d_2 z^2 + d_3z^3,
 \label{DcPolin}
\end{equation}
where $d_2$ and $d_3$ are free parameters. From Eq.(\ref{dCEz}), we may write
\begin{equation}
 E(z)=\bigg[\frac{dD_C(z)}{dz}\bigg]^{-1}.
 \label{EzDc}
\end{equation}

Naturally, from Eqs.(\ref{EzDc}) and (\ref{DcPolin}), one obtains
\begin{equation}
E(z)=\frac{1}{1 + 2d_2 z + 3d_3 z^2}\,\, .
\label{Ezdcpolin}
\end{equation}

Finally, from Eqs. (\ref{DL}), (\ref{DcPolin}) and (\ref{dMsinn}) the dimensionless luminosity distance is
\begin{equation}
\label{dl2}
D_L(z)= (1+z)\sinn(z + d_2 z^2 + d_3z^3,\Omega_k)\,.
\end{equation}

Equations (\ref{Ezdcpolin}) and (\ref{dl2}) shall be compared with $H(z)$ measurements and luminosity distances from SNe Ia, respectively, in order to determine $d_2$ and $\Omega_k$.

\subsection{\texpdf{$\Omega_k$}{Wk} from \texpdf{$H(z)$}{Hz}}
In order to assess $\Omega_k$ by means of $H(z)$ we need an expression for $H(z)$. If one wants to avoid dynamical assumptions, one must resort to kinematical methods which use an expansion of $H(z)$ over the redshift.

Let us try a simple $H(z)$ expansion, namely, the quadratic expansion:
\begin{equation}
 \frac{H(z)}{H_0} =E(z)= 1 + h_1z + h_2z^2.
 \label{EzPolin}
\end{equation}

In order to constrain the model with SNe Ia data, we obtain the luminosity distance from Eqs.(\ref{DL}), (\ref{Dc}) and (\ref{EzPolin}). We have
\begin{equation}
 D_C=\int_0^z\frac{dz'}{E(z')}=\int_0^z\frac{dz'}{1 + h_1z' + h_2z'^2},
\end{equation}
which gives three possible solutions, according to the sign of $\Delta\equiv h_1^2-4h_2$, such as
\begin{eqnarray}
 D_C=\left\{\begin{array}{ll}
\dfrac{2}{\sqrt{-\Delta}}\left[\atan\left(\dfrac{2h_2z+h_1}{\sqrt{-\Delta}}\right)-\atan\dfrac{h_1}{\sqrt{-\Delta}}\right],&\Delta<0,\\
\dfrac{2z}{h_1z+2},&\Delta=0,\\
\dfrac{1}{\sqrt{\Delta}}\ln\left|\left(\dfrac{\sqrt{\Delta}+h_1}{\sqrt{\Delta}-h_1}\right)\left(\dfrac{\sqrt{\Delta}-h_1-2h_2z}{\sqrt{\Delta}+h_1+2h_2z}\right)\right|,&\Delta>0,
\end{array}
 \right.\label{eqDCz}
\end{eqnarray}
from which follows the luminosity distance $D_L(z)=(1+z)\sinn(D_C,\Omega_k)$.

\begin{figure*}
\centering
\includegraphics[width=.49\linewidth]{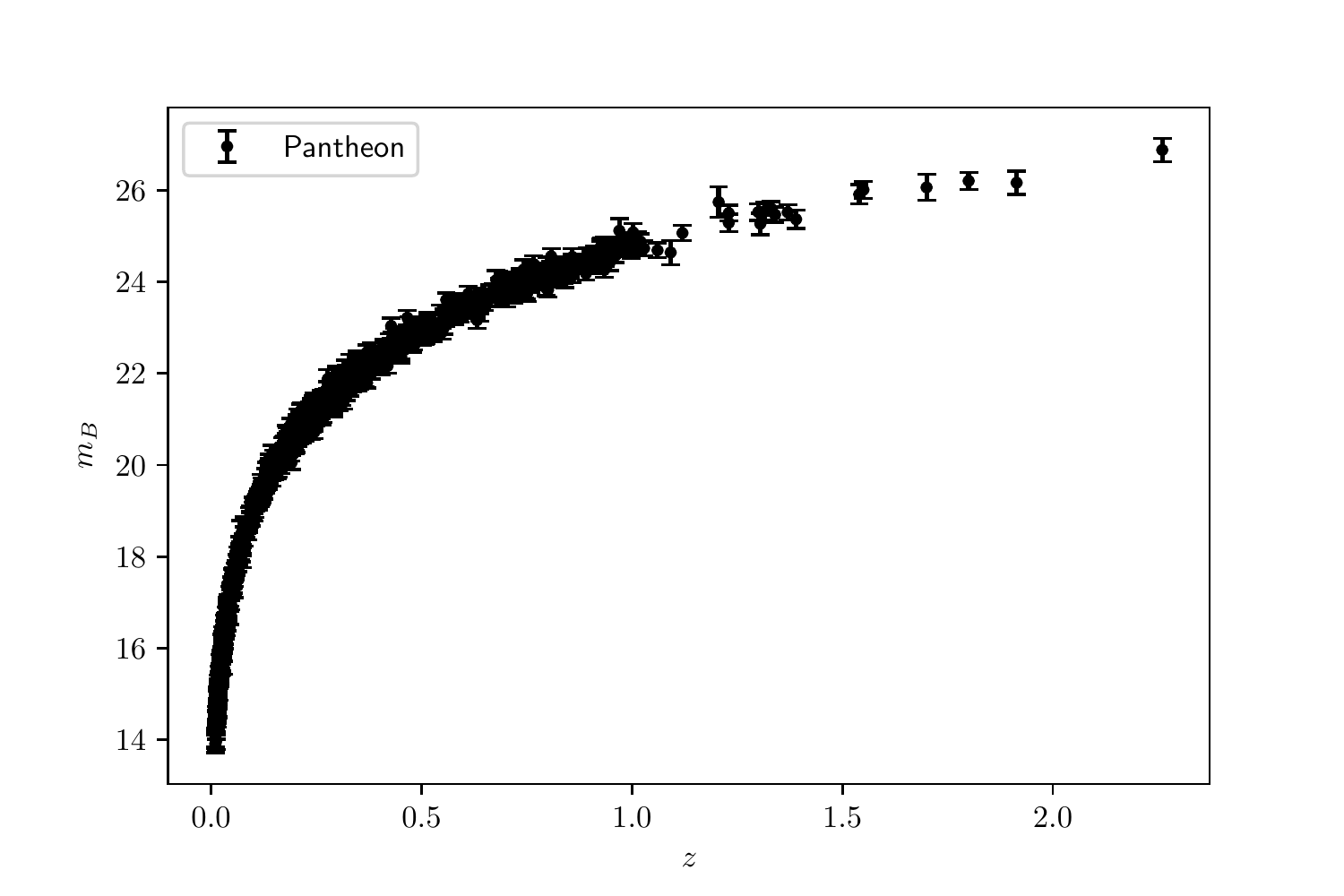}
 \includegraphics[width=0.49\linewidth]{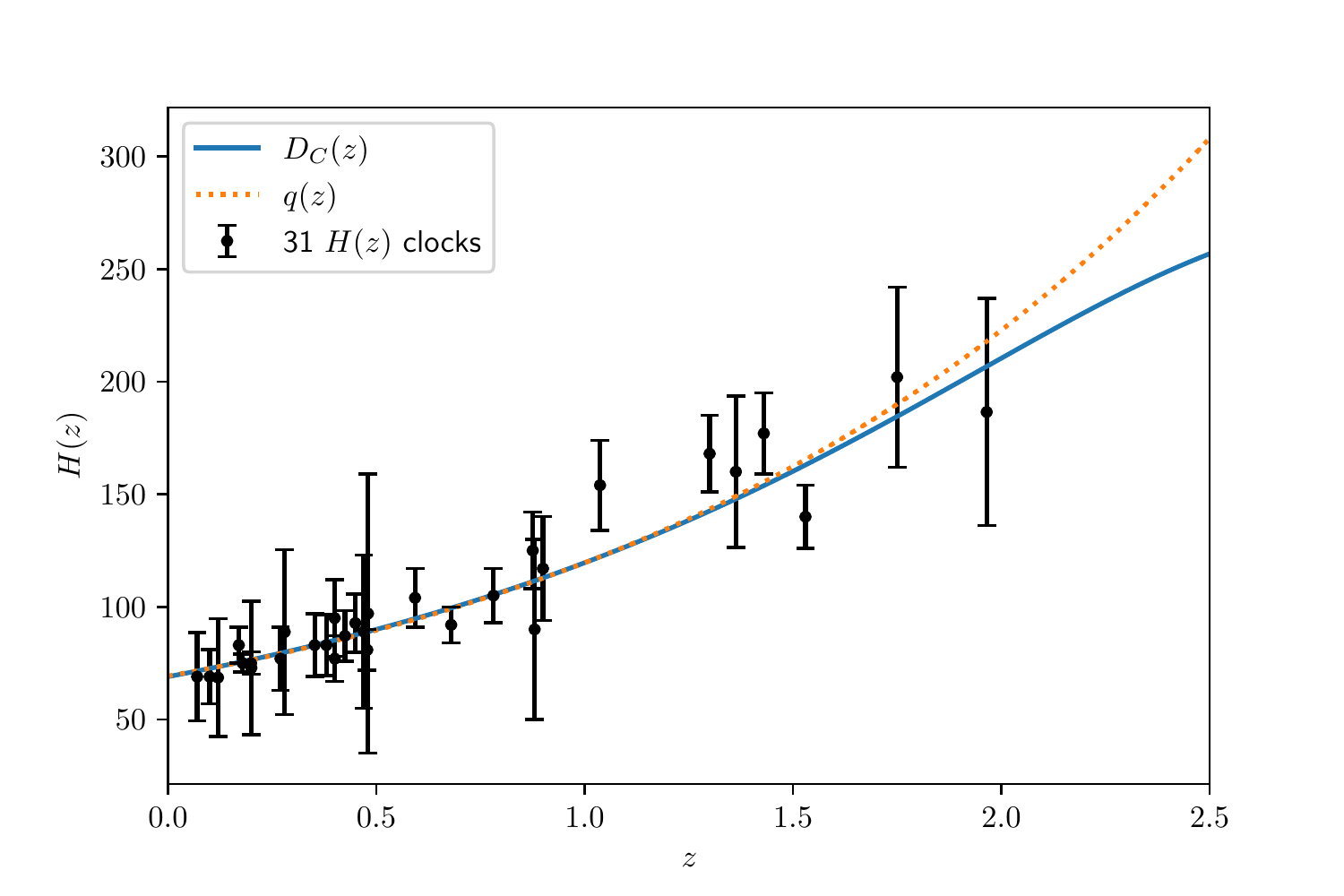}
 \caption{\label{Data} {\bf a)} SNe Ia apparent magnitude $m_B$ from Pantheon. The error bars shown correspond only to statistical errors, but we use the full covariance matrix (statistical+systematic errors) in the analysis. {\bf b)} 31 $H(z)$ cosmic chronometers. The lines represent the best fit from SNe+$H(z)$ data for each model.
 }
 \label{AllData}
\end{figure*}

\subsection{\texpdf{$\Omega_k$}{Wk} from \texpdf{$q(z)$}{qz}}
Now we can analyze $\Omega_k$ by parametrizing $q(z)$.  From (\ref{qzH}) one may find $E(z)$ as
\begin{equation}
 E(z)=\exp\left[\int_0^z\frac{1+q(z')}{1+z'}dz'\right].
\end{equation}
If we assume a linear $z$ dependence in $q(z)$, as
\begin{equation}
 q(z)=q_0+q_1z,\label{eqqz}
\end{equation}
which is the simplest $q(z)$ parametrization that allows for an acceleration transition 
as required by SNe Ia data \citep{RiessEtAl04,zt}, one may find
\begin{equation}
 E(z)=e^{q_1z}(1+z)^{1+q_0-q_1},\label{eqEzq}
\end{equation}
while the line-of-sight comoving distance $D_C(z)$ (\ref{Dc}) is given by
\begin{equation}
 D_C(z)=e^{q_1}q_1^{q_0-q_1}\left[\Gamma(q_1-q_0,q_1)-\Gamma(q_1-q_0,q_1(1+z))\right],
\end{equation}
where $\Gamma(a,x)$ is the incomplete gamma function defined in \citep{AbraSteg} as $\Gamma(a,x)\equiv\int_x^\infty e^{-t}t^{a-1}dt$, with $a>0$, from which follows the luminosity distance as $D_L(z)=(1+z)\sinn(D_C,\Omega_k)$, which can be constrained from observational data.

\section{Samples}\label{samples}
\subsection{\texpdf{$H(z)$}{Hz} dataset}
In order to constrain the free parameters, we use the Hubble parameter ($H(z)$) data in different redshift values. These kind of observational data are quite reliable because in general such observational data are independent of the background cosmological model, just relying on astrophysical assumptions. We have used the currently most complete compilation of $H(z)$ data, with 51 measurements \citep{MaganaEtAl18}.

At the present time, the most important methods for obtaining $H(z)$ data are\footnote{See \citep{zt} for a review.} (i) through ``cosmic chronometers'', for example, the differential age of galaxies (DAG) \citep{Simon05,Stern10,Moresco12,Zhang12,Moresco15,MorescoEtAl16}, (ii) measurements of peaks of acoustic oscillations of baryons (BAO) \citep{Gazta09,Blake12,Busca12,AndersonEtAl13,Font-Ribera13,Delubac14} and (iii) through correlation function of luminous red galaxies (LRG) \citep{Chuang13,Oka14}.

Among these methods for estimating $H(z)$, the 51 data compilation as grouped by \citep{MaganaEtAl18}, consists of 20 clustering (BAO+LRG) and 31 differential age $H(z)$ data.

Differently from \citep{MaganaEtAl18}, we choose not to use $H_0$ in our main results here, due to the current tension among $H_0$ values estimated from different observations \citep{RiessEtAl16,Planck16,BernalEtAl16}.

The method used to estimate $H(z)$ data from BAO depends on the choice of a fiducial cosmological model. Even if it has an weak model dependence, we choose here not to work with the $H(z)$ data from BAO. So, in order to keep the analysis the most model-independent possible, we shall work here only with the 31 differential age $H(z)$ data (cosmic chronometers) from \citep{MaganaEtAl18}.

\subsection{SNe Ia}
We have chosen to work with one of the largest SNe Ia sample to date, namely, the Pantheon sample \citep{pantheon}. This sample consists of 279 SNe Ia from Pan-STARRS1 (PS1) Medium Deep Survey ($0.03<z<0.68$), combined with distance estimates of SNe Ia from Sloan Digital Sky Survey (SDSS), SNLS and various low-$z$ and Hubble Space Telescope samples to form the largest combined sample of SNe Ia, consisting of a total of 1048 SNe Ia in the range of $0.01<z<2.3$.

As explained on \citep{pantheon}, the PS1 light-curve fitting has been made with SALT2 \citep{GuyEtAl10}, as it has been trained on the JLA sample \citep{BetouleEtAl14}. 
Three quantities are determined in the light-curve fit that are needed to derive a distance: the colour $c$, the light-curve shape parameter $x_1$ and the log of the overall flux normalization $m_B$. We can see the $m_B$ data for Pantheon at Fig. \ref{AllData}a.

The SALT2 light-curve fit parameters are transformed into distances using a modified version of the Tripp formula \citep{Tripp98},
\begin{equation}
 \mu=m_B-M+\alpha x_1-\beta c+\Delta_M+\Delta_B,
 \label{mupanth}
\end{equation}
where $\mu$ is the distance modulus, $\Delta_M$ is a distance correction term based on the host galaxy mass of the SN, and $\Delta_B$ is a distance correction factor based on predicted biases from simulations. As can be seen, $\alpha$ is the coefficient of the relation between luminosity and stretch, while $\beta$ is the coefficient of the relation between luminosity and color, and $M$ is the absolute $B$-band magnitude of a fiducial SN Ia with $x_1=0$ and $c=0$.

Differently from previous SNe Ia samples, like JLA \citep{BetouleEtAl14}, Pantheon uses a calibration method named BEAMS with Bias Corrections (BBC), which uses cosmological simulations assuming  a  reference  $\Lambda$CDM  model. The  cosmological  dependence  is expected  to  be  small, so neglecting this dependence, allows one to determine SNe Ia distances without one having to fit SNe parameters jointly with cosmological parameters. Thus, Pantheon provide directly corrected $m_B$ estimates in order for one to constrain cosmological parameters alone.

The systematic uncertainties were propagated through a systematic uncertainty matrix. An
uncertainty matrix C was defined such that
\be\label{cov}
\mathbf{C} = \mathbf{D}_{\mathrm{stat}} + \mathbf{C}_{\mathrm{sys}}.
\ee

The statistical matrix $\mathbf{D}_\mathrm{stat}$ has only a diagonal component that includes photometric errors of the SN distance, the distance uncertainty from the mass step correction, the uncertainty from the distance bias correction, the uncertainty from the peculiar velocity uncertainty and redshift measurement uncertainty in quadrature, the uncertainty from stochastic gravitational lensing, and the intrinsic scatter.

\section{Analyses and Results}\label{analysis}
In our analyses, we have chosen flat priors for all parameters, so always the posterior distributions are proportional to the likelihoods.

For $H(z)$ data, the likelihood distribution function is given by $\like_H \propto e^{-\frac{\chi^2_H}{2}}$,
\\
\\
with 
\begin{equation}
\chi^2_H = \sum_{i = 1}^{31}\frac{{\left[ H_{obs,i} - H(z_i,\mathbf{s})\right] }^{2}}{\sigma^{2}_{H_i,obs}}\,.
\label{chi2H}
\end{equation}
\\
The $\chi^2$ function for Pantheon is given by
\be
\chi^2_{\text{SN}}=\mathbf{\Delta m}^T\cdot\mathbf{C}^{-1}\cdot\mathbf{\Delta m},
\ee
where $\textbf{C}$ is the same from \eqref{cov}, $\mathbf{\Delta m}=m_B-m_{\mathrm{mod}}$, and
\be
m_{\mathrm{mod}}=5\log_{10}D_L(z)+\mathcal{M}\,,
\ee
where $\mathcal{M}$ is a nuisance parameter which encompasses $H_0$ and $M$. We choose to project over $\mathcal{M}$, which is equivalent to marginalize the likelihood $\like_\text{SN}\propto e^{-\chi^2_\text{SN}/2}$ over $\mathcal{M}$, up to a normalization constant. In this case we find the projected $\chi^2_{proj}$:
\be
\chi^2_{proj}=S_{mm}-\frac{S_m^2}{S_A}\,,
\ee
where $S_{mm}=\sum_{i,j}\Delta m_i\Delta m_jA_{ij}=\mathbf{\Delta m}^T\cdot\mathbf{A}\cdot\mathbf{\Delta m}$, $S_m=\sum_{i,j}\Delta m_iA_{ij}=\mathbf{\Delta m}^T\cdot\mathbf{A}\cdot\mathbf{1}$, $S_A=\sum_{i,j}A_{ij}=\mathbf{1}^T\cdot\mathbf{A}\cdot\mathbf{1}$ and $\mathbf{A}\equiv\mathbf{C}^{-1}$.

\begin{figure}
 \includegraphics[width=\linewidth]{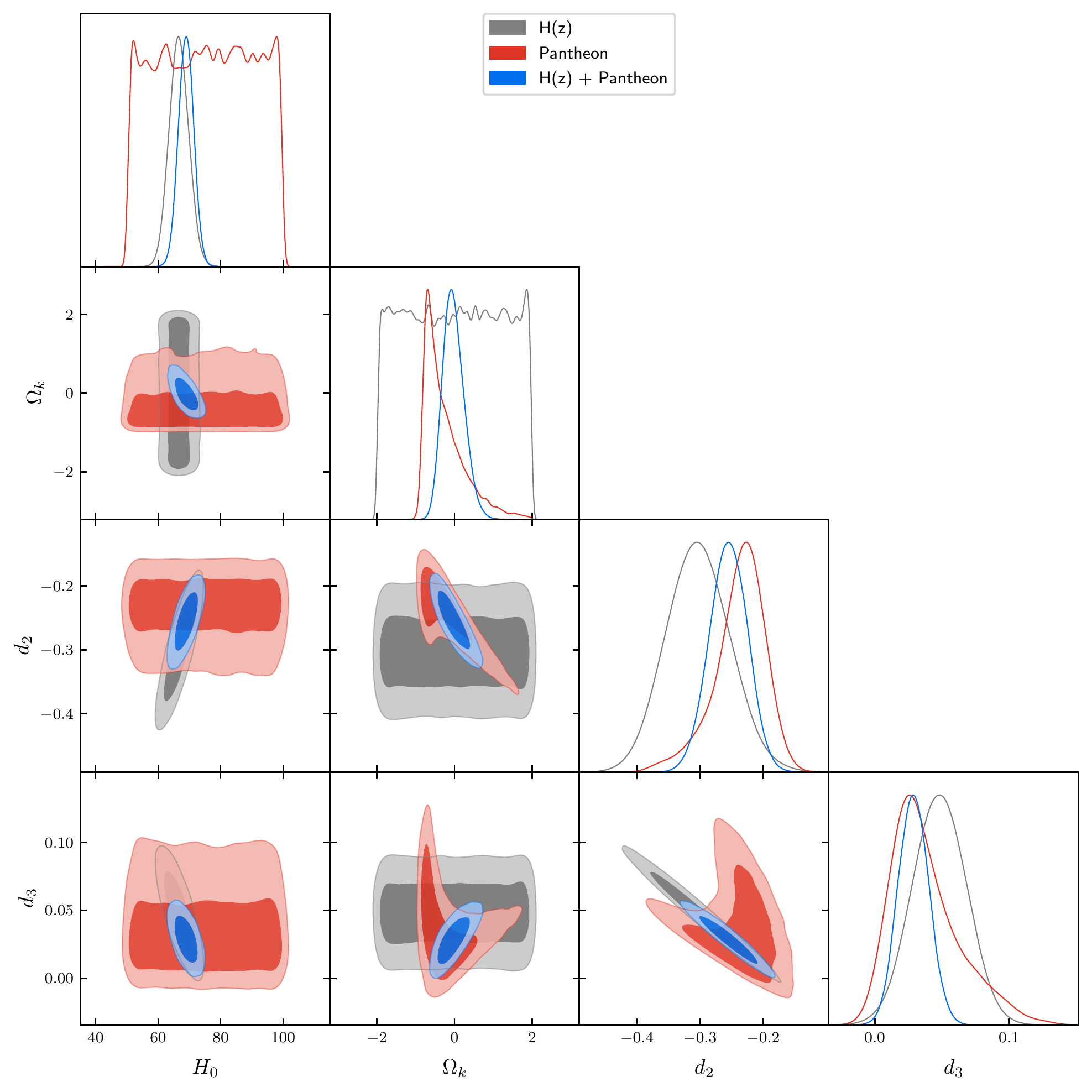}
 \caption{\label{DcPolinTriangAll} All constraints from Pantheon and cosmic chronometers for $D_C(z)=z+d_2z^2+d_3z^3$. The contours correspond to 68.3\% c.l. and 95.4\% c.l.}
\end{figure}

\begin{figure}
 \includegraphics[width=\linewidth]{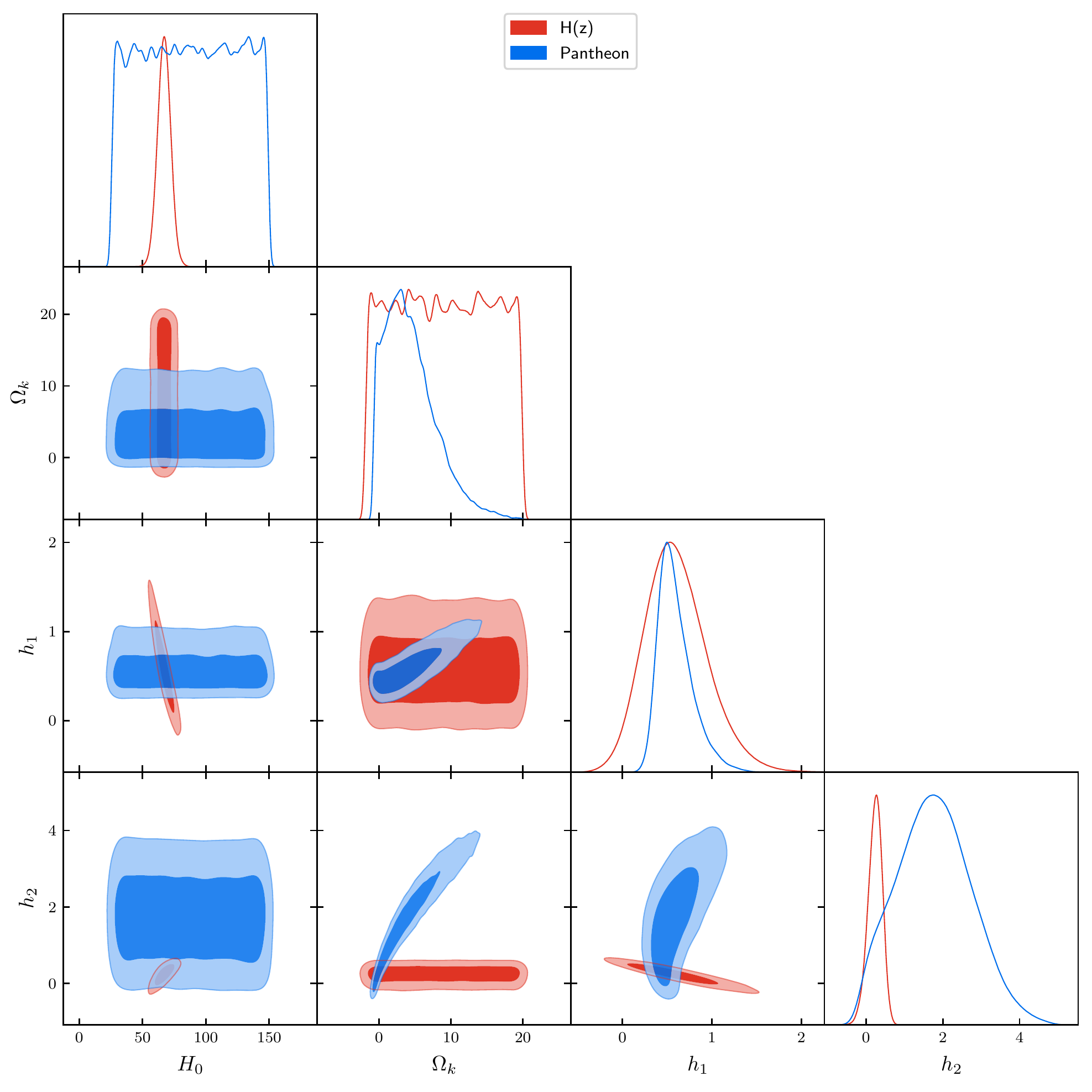}
 \caption{\label{HzPolinTriangAll} All constraints from Pantheon and cosmic chronometers for $H(z)=H_0(1+h_1z+h_2z^2)$. The contours correspond to 68.3\% c.l. and 95.4\% c.l.}
\end{figure}

\begin{figure}
 \includegraphics[width=\linewidth]{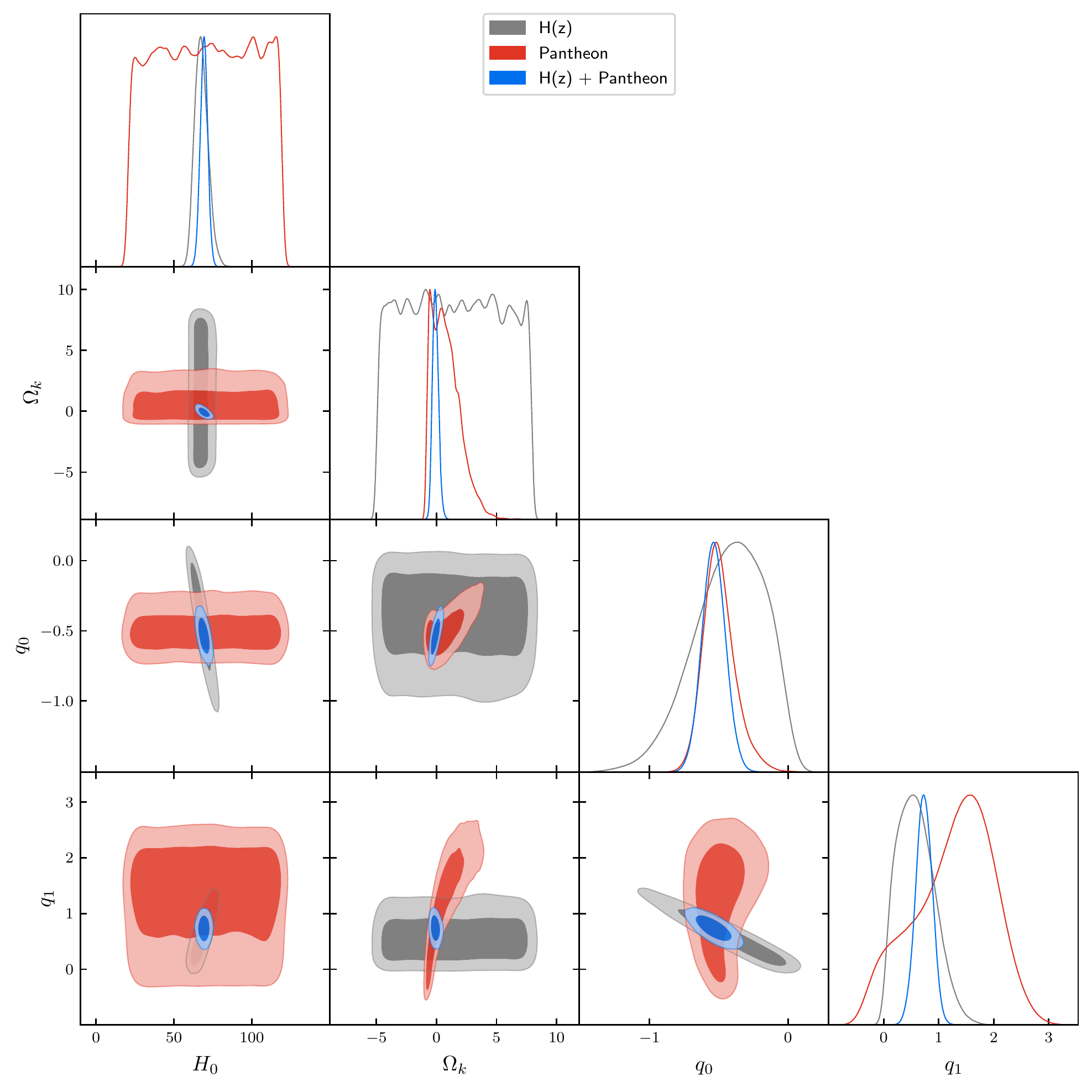}
 \caption{\label{q0q1TriangAll} All constraints from Pantheon and cosmic chronometers for $q(z)=q_0+q_1z$. The contours correspond to 68.3\% c.l. and 95.4\% c.l.}
\end{figure}

\begin{figure}
 \includegraphics[width=\linewidth]{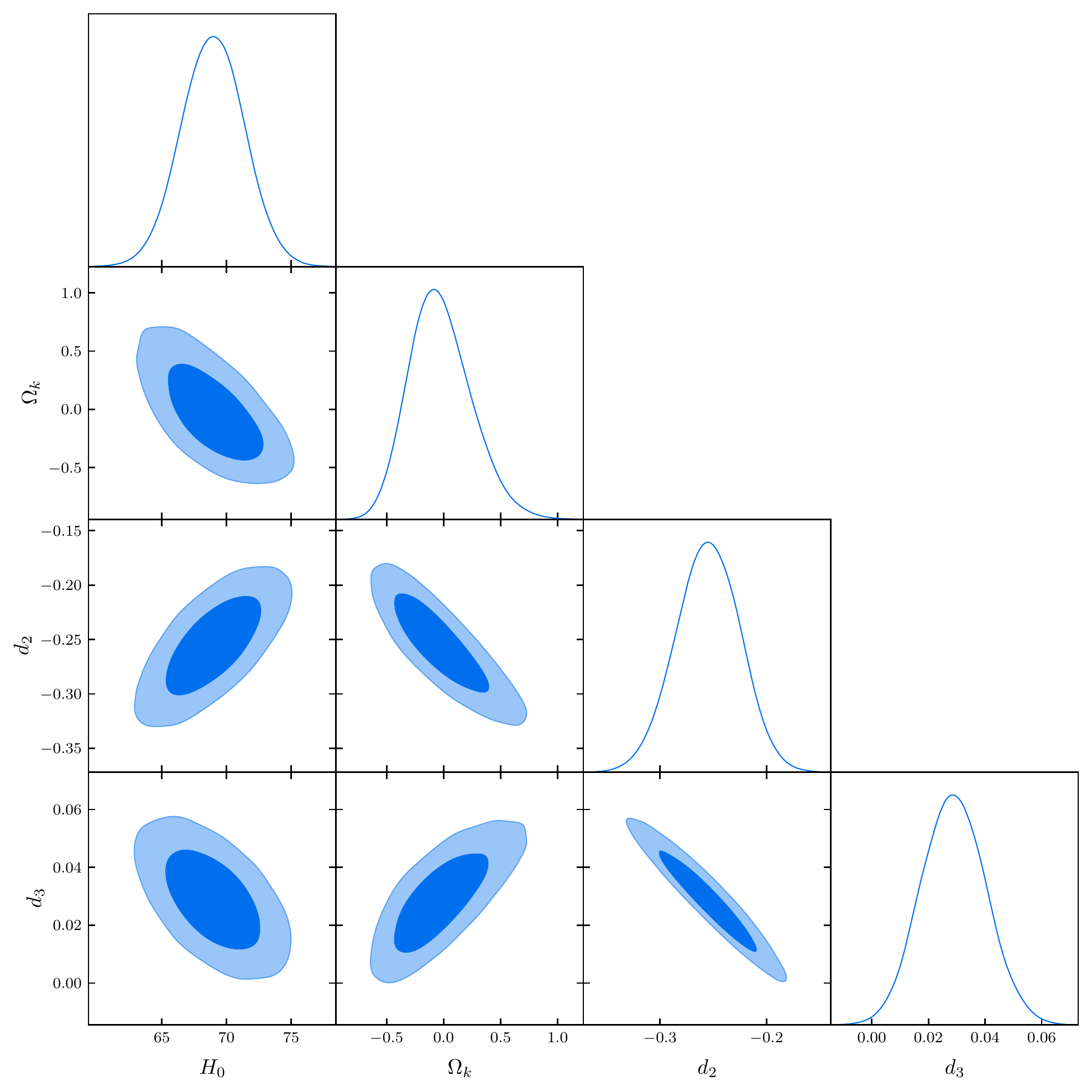}
 \caption{\label{DcPolinTriang} Combined constraints from Pantheon and $H(z)$ for $D_C(z)=z+d_2z^2+d_3z^3$. The contours correspond to 68.3\% c.l. and 95.4\% c.l.}
\end{figure}



\begin{figure}
 \includegraphics[width=\linewidth]{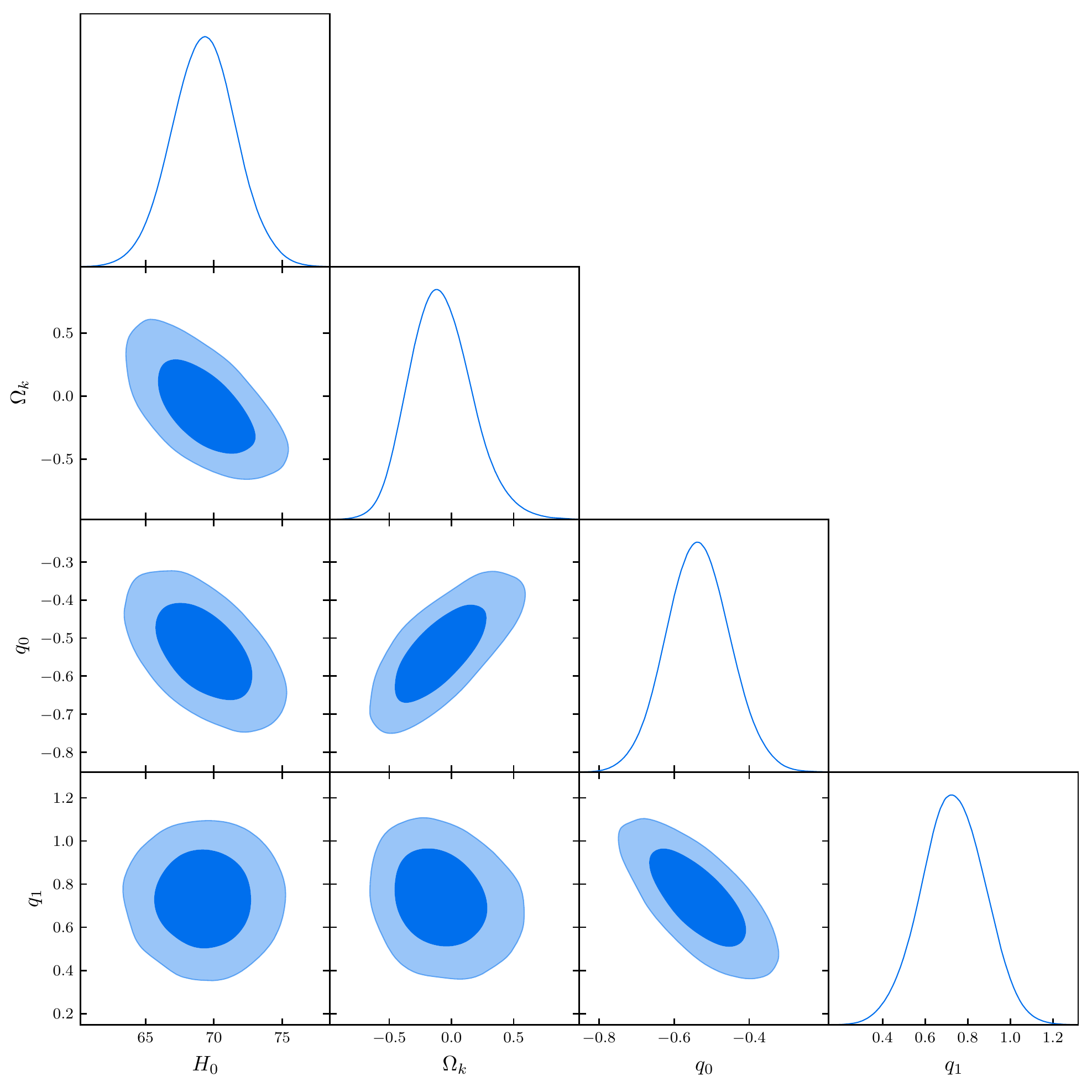}
 \caption{\label{q0q1Triang} Combined constraints from Pantheon and $H(z)$ for $q(z)=q_0+q_1z$. The contours correspond to 68.3\% c.l. and 95.4\% c.l.}
\end{figure}

In order to obtain the constraints over the free parameters,  the likelihood $\like\propto e^{-\chi^2/2}$,  where $\chi^2\equiv\chi^2_H+\chi^2_{proj}$, has been sampled through a Monte Carlo Markov Chain (MCMC) analysis. A simple and powerful MCMC method is the so called Affine Invariant MCMC Ensemble Sampler by \citep{GoodWeare}, which was implemented in {\sffamily Python} language with the {\sffamily emcee} software by \citep{ForemanMackey13}.

\begin{figure}
\centering
 \includegraphics[width=\linewidth]{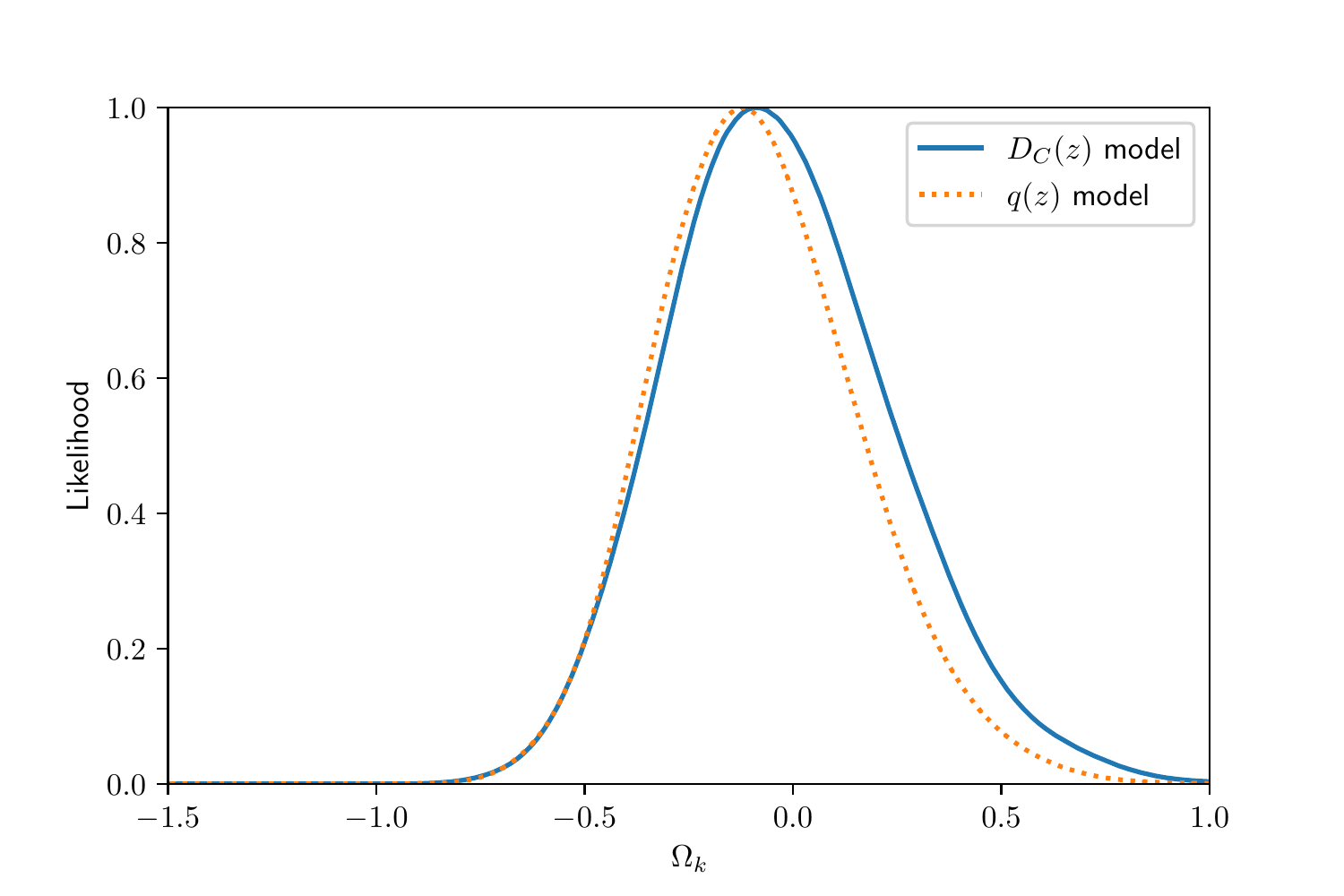}
 \caption{\label{wklikes} Likelihoods for spatial curvature density parameter from Pantheon and $H(z)$ data combined. Blue solid line corresponds to $D_C(z)$ parametrization, orange long-dashed line corresponds to $q(z)$ parametrization. $H(z)$ parametrization was not combined, as mentioned in the text.}      
\end{figure}

We used the free software {\sffamily emcee} to sample from our likelihood in $n$-dimensional parameter space.
In order to plot all the constraints on each model in the same figure, we have used the freely available software {\sffamily getdist}\footnote{{\sffamily getdist} is part of the great MCMC sampler and CMB power spectrum solver {\sffamily COSMOMC}, by \citep{cosmomc}.}, in its {\sffamily Python} version. The results of our statistical analyses can be seen on Figs. \ref{DcPolinTriangAll}-\ref{wklikes} and on Table \ref{tab1}.

In Figs. \ref{DcPolinTriangAll}-\ref{q0q1TriangAll}, we show explicitly the independent constraints, in order to see the complementarity between SNe Ia and $H(z)$ data. First of all, as expected, SNe Ia does not constrain $H_0$. In SNe confidence level contours, $H_0$ is only limited by our prior, but $H(z)$ data gives good constraints over $H_0$. We can see also, that in general, SNe Ia alone does not constrain well $\Omega_k$, but by combining with $H(z)$, which constrain the other parameters, good constraints over the curvature are found. In the planes not containing $\Omega_k$ ($d_2-d_3$, $h_1-h_2$ and $q_0-q_1$) we can see that $H(z)$ also helps to reduce a lot the allowed parameter space.

In Figs. \ref{DcPolinTriang}-\ref{q0q1Triang}, we have the combined results for each parameterization, where we can clearly see how the combination SNe Ia+$H(z)$ yields good constraints over $\Omega_k$, as well as the other kinematic parameters. For all parametrizations, the best constraints over the spatial curvature comes from $q(z)$ model, as can be seen on Fig. \ref{wklikes}. We can also see in this Figure that all constraints are compatible at 1$\sigma$ c.l. Finally, Table \ref{tab1} shows the full numerical results from our statistical analysis. 

\begin{table}
\centering
\begin{tabular}{|l|c|c|}
\hline
 Parameter &  $D_C(z)$                           & $q(z)$                           \\
\hline
$H_0$      & $69.0\pm 2.4\pm4.9        $         & $69.3\pm 2.4^{+4.8}_{-4.7}      $\\
$\Omega_k$ & $-0.03^{+0.24+0.56}_{-0.30-0.53}  $ & $-0.08^{+0.21+0.54}_{-0.27-0.45}$\\
$d_2$      & $-0.255\pm 0.030^{+0.059}_{-0.061}$ & --                               \\
$d_3     $ & $0.029\pm0.011^{+0.023}_{-0.022}$   & --                               \\
$q_0     $ & --                                  & $-0.536\pm0.085\pm0.17$          \\
$q_1$      & --                                  & $0.73\pm0.15\pm0.30$             \\
\hline
\end{tabular}
\caption{Constraints from Pantheon+$H(z)$ for $D_C(z)$ and $q(z)$ parametrizations. The central values correspond to the mean and the 1 $\sigma$ and 2 $\sigma$ c.l. correspond to the minimal 68.3\% and 95.4\% confidence intervals.}
\label{tab1}
\end{table}

\newpage

Comparing with previous results in the literature, \citet{LiEtAl16} have combined 22 $H(z)$ data from cosmic chronometers with Union 2.1 SNe Ia data and JLA SNe Ia data. The combination with Union 2.1 yielded $\Omega_k=-0.045^{+0.176}_{-0.172}$ and they found $\Omega_k=-0.140^{+0.161}_{-0.158}$ from JLA combination. \citet{WangEtAl17} have put model independent constraints over $\Omega_k$ and opacity from JLA SNe Ia data and 30 $H(z)$ data. They have used Gaussian Processes method and have obtained $\Omega_k=0.44\pm0.64$, with a high uncertainty, due to degeneracy with opacity.  It is worth to mention that, although model-independent, both \citep{LiEtAl16} and \citep{WangEtAl17} have followed a different approach from the present paper. They do not parametrize any cosmological observable, instead they obtain a distance modulus from $H(z)$ data, and compare with distance modulus from SNe Ia, which are dependent on spatial curvature. As already mentioned, \citet{YuEtAl18} have used $H(z)$ and BAO, with the aid of Gaussian Processes and have found $\Omega_k=-0.03\pm0.21$, consistent with our results. By combining CMB data with BAO, in the context of $\Lambda$CDM, the \citet{Planck18} have found $\Omega_k=0.001\pm0.002$. It is consistent with our result, but it is dependent on the chosen dynamical model, $\Lambda$CDM.

Another interesting result that can be seen on Table \ref{tab1} is the $H_0$ constraint. As one may see, the constraints over $H_0$ are consistent among both parametrizations. The constraints over $H_0$ are quite stringent today from many observations \citep{RiessEtAl19,Planck18}. However, there is some tension among $H_0$ values estimated from Cepheids \citep{RiessEtAl19} and from CMB \citep{Planck18}. While Riess {\it et al.} advocate $H_0=74.03\pm1.42$ km/s/Mpc, the Planck collaboration analysis, in the context of $\Lambda$CDM, yields $H_0=67.4\pm0.5$ km/s/Mpc, a 4.4$\sigma$ lower value. 

It is interesting to note, from our Table \ref{tab1} that, although we are working with model independent parametrizations and data at intermediate redshifts, our result is in better agreement with the high redshift result from Planck. In fact, all our results are compatible within 1$\sigma$ with the Planck's result, while, for the Riess' result, our $D_C(z)$ result is marginally compatible at 1.8$\sigma$, and $q(z)$ is marginally compatible at 1.7$\sigma$.



\section{Conclusion}\label{conclusion}

In the present work, we wrote the comoving distance $ D_C $,  the Hubble parameter $H (z) $ and  the deceleration parameter $q(z)$ as  third, second and first degree polynomials on $z$, respectively (see equations (\ref{DcPolin}), (\ref{EzPolin}) and (\ref{eqqz})), and obtained, for each case, the $\Omega_k$ value. We have shown that by combining Supernovae type Ia data and Hubble parameter measurements, nice constraints are found over the spatial curvature, without the need of assuming any particular dynamical model. Our results can be found in Figures \ref{DcPolinTriangAll}-\ref{q0q1Triang}. As one may see from Figs. \ref{DcPolinTriangAll}-\ref{q0q1TriangAll}, the analyses by using SNe Ia and $H(z)$ data  are complementary to each other, providing tight limits in the parameter spaces. As a result, the values obtained for the spatial curvature in each case were $\Omega_k=-0.03^{+0.24}_{-0.30}$ and  $-0.08^{+0.21}_{-0.27}$ at 1$\sigma$ c.l., for $D_C(z)$ and $q(z)$ parametrizations (see Fig. \ref{wklikes}), all compatible with a spatially flat Universe, as predicted by most inflation models and confirmed by CMB data, in the context of $\Lambda$CDM model. The $H(z)$ parametrization presented incompatibilities from its constraints coming from SNe Ia and cosmic clocks data and was not considered in the joint analysis.

Further investigations could include different parametrizations and other kinematical methods in order to determine the Universe spatial curvature independently from the matter-energy content.


\section{\label{acknowledgements}Acknowledgements}
JFJ is supported by Funda\c{c}\~ao de Amparo \`a Pesquisa do Estado de S\~ao Paulo - FAPESP (Process no. 2017/05859-0). RV and MM are supported by  Funda\c{c}\~ao de Amparo \`a Pesquisa do Estado de S\~ao Paulo - FAPESP (thematic project process no. 2013/26258-2 and regular project process no. 2016/09831-0). MM is also supported by CNPq and Capes. PHRSM also thanks CAPES for financial support.

\section*{Data Availability}
No new data were generated or analysed in support of this research.



\bibliographystyle{mnras}
\bibliography{example} 

\bsp	
\label{lastpage}
\end{document}